\documentclass[preprint,12pt]{revtex4}
 
\usepackage{amsmath}
 
\usepackage{setspace}
  
\usepackage[dvipsnames,rgb,dvips]{xcolor}

\usepackage{graphicx,mathrsfs,psfrag}
\usepackage{dcolumn}

\newcommand{\beq}{\begin{equation}}
\newcommand{\eeq}{\end{equation}}

\newcommand{\ud}{\mathrm{d}}

\newcommand{\lapu}{ \boldsymbol{\Delta}{\textbf{u}}}
\newcommand{\divu}{{\boldsymbol{\nabla}}.{\textbf{u}}}

\newcommand{\upp}{{\textbf{u}}^{n+1}}
\newcommand{\upn}{{\textbf{u}}^{n}}
\newcommand{\ustar}{{\textbf{u}}^{*}}

\newcommand{\lpp}{{\boldsymbol{\lambda}}^{n+1}}
\newcommand{\lp}{{\boldsymbol{\lambda}}^{n}}

\newcommand{\disct}{\Delta t}
\newcommand{\rr}{\rho _r}
\newcommand{\rouge}{}

\begin{document}
  
\title {Chaotic sedimentation of particle pairs in a vertical channel at low Reynolds number: multiple states and routes to chaos.} 
  
\author{Romuald Verjus$^{(1)}$, Sylvain Guillou$^{(1)}$, Alexander Ezersky$^{(2)}$ and Jean-R\'egis Angilella$^{(1)}$\footnote{Corresponding author: Jean-Regis.Angilella@unicaen.fr}}
%\thanks{}
\affiliation{\mbox{}$^{(1)}$ Universit\'e de Caen Basse Normandie, LUSAC, Cherbourg, France}
 
\affiliation{\mbox{}$^{(2)}$ Universit\'e de Caen Basse Normandie, M2C, Caen, France}
 
\begin{abstract}

  The sedimentation of a pair of rigid circular particles in a two-dimensional vertical channel containing a Newtonian fluid is investigated numerically, for terminal particle Reynolds numbers ($\mbox{Re}_T$) ranging from 1 to  10, and  for a confinement ratio equal to 4. While it is widely admitted that sufficiently inertial pairs should sediment by performing a regular DKT oscillation (Drafting-Kissing-Tumbling), the present analysis shows in contrast that a chaotic regime can also exist for such particles, leading to a much slower sedimentation velocity. It consists of a nearly horizontal pair, corresponding to a maximum effective blockage ratio, and performing a quasiperiodic transition to chaos under increasing the particle weight. For less inertial regimes, the classical oblique doublet structure and its complex behavior (multiple stable states and hysteresis, period-doubling cascade and chaotic attractor) are recovered, in agreement with previous work [Aidun \& Ding, Physics of Fluids 15(6), 2003].  As a consequence of these various behaviors, the link between the terminal Reynolds number and the non-dimensional driving force is complex: it contains several branches displaying hysteresis as well as various bifurcations. For the range of Reynolds number considered here, a global bifurcation diagram is given.

\end{abstract}

 \maketitle

\section{Introduction}

\begin{figure}[h] 
\center{\includegraphics[scale=0.6]{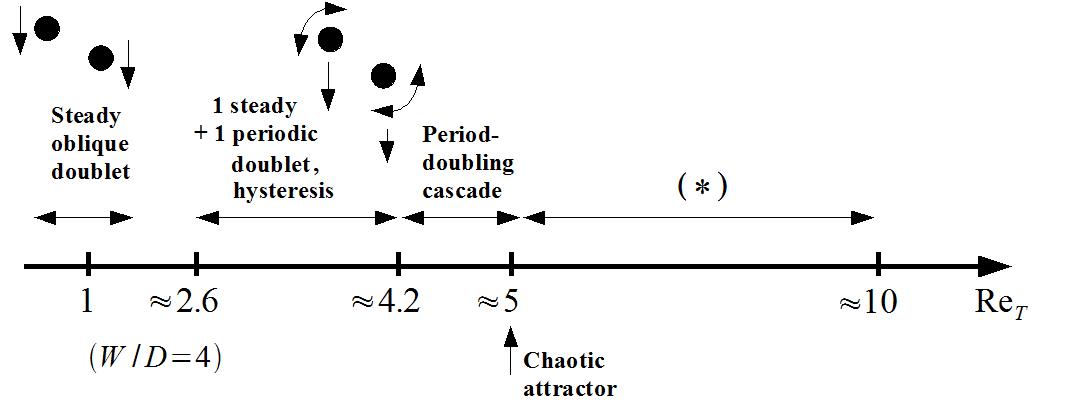}}
\caption{Sketch of the typical behaviors of the sedimenting pair described in the literature, when the confinement ratio is equal to 4. The star $(\ast)$ indicates the range of Reynolds numbers explored in the present study.}
\label{TypicalBehaviors}
\end{figure}

The sedimentation of inertial particles at low Reynolds number has been intensively studied in the past, since this problem is ubiquitous in industrial or natural sciences \citep{Richardson1954,Jayaweera1965,Crowe1998,Derksen2011,Feng1996,Champmartin2006}.
 Even if particles are non-Brownian and  not submitted to electrostatic forces, they strongly interact 
in general through hydrodynamic interactions. This leads to complex sedimentation regimes with very irregular
individual trajectories.  
\rouge{A simple though non-trivial situation related to this problem is the settling of a small number of spheres in a vertical channel: the settling velocity, being influenced by inter-particle as well as particle/wall hydrodynamic interactions, is difficult to predict, especially if the confinement is strong. In the context of fluidized bed analyses, Fortes {\it et al.} \cite{Fortes1987} observed that pairs of spheres, settling in a rectangular channel with a thin gap, could have a complex behavior. For Reynolds numbers of a few hundred, these authors observed a Drafting-Kissing-Tumbling (DKT) phenomenon: the trailing sphere approaches the leading one, then overtakes it and becomes the leading sphere, and so on. These experiments also revealed that spheres could place themselves along a quasi horizontal line joining the two end walls of the channel, and slowly sediment in this stable position.
The two-dimensional version of this problem is the settling of {\it disks} in a vertical plane channel. It has been studied numerically by Feng {\it et al.} \cite{Feng1995}, Aidun \& Ding \cite{Aidun2003}, and more recently by 
Wang {\it et al.} \cite{Wang2014}. Even though the detailed structure of the flow around disks differs from the case of spheres, some common features exist between the two situations.  In particular, the DKT phenomenon has been observed for disks also \cite{Feng1995} \cite{Aidun2003}. For less inertial regimes, the  pair converges to some steady sedimentation structure taking the form of an oblique doublet (as shown in the pioneering works by Feng {\it et al.} \cite{Feng1995}).}  Because these behaviors occur for a wide range of Reynolds numbers, it is widely believed that the oblique doublet and the DKT are the only possible configurations for this sedimentation problem. However, Aidun \& Ding \cite{Aidun2003} revealed that the motion of the particle pair could be much more complex. By using a Lattice Boltzmann approach, they observed multiple stable states, hysteresis, as well as a period-doubling cascade leading to a chaotic attractor.
The steady oblique doublet exists when the terminal Reynolds number, based on the particle diameter $D$, the long-term sedimentation velocity $V_T$ (averaged over time and over both particles), and the fluid kinematic viscosity $\nu$:
\beq
\mbox{Re}_{T} = \frac{V_T \, D}{\nu},
\eeq
is below some value of order unity. 
When $\mbox{Re}_T$ increases  the oblique doublet  bifurcates, leading to a variety of behaviors as sketched in Fig.\ \ref{TypicalBehaviors}.  For $\mbox{Re}_T$ between 2.6 and 4.2, two  sedimentation structures exist, corresponding to two different terminal velocities. Both structures are observed to oscillate periodically around an oblique line while the particles sediment. However, the slowest configuration, corresponding to a more horizontal doublet, has been observed to become unstable \cite{Aidun2003} when $\mbox{Re}_T$ is above 4.2: for such Reynolds numbers a single, time-periodic, sedimentation structure is observed. By further increasing the Reynolds number a dramatic change is observed, as the doublet undergoes a period-doubling cascade leading to a low-dimensional chaotic attractor when $\mbox{Re}_T$ is close to 5. This chaotic dynamics vanishes however for larger Reynolds numbers and a periodic Drafting-Kissing-Tumbling phenomenon takes place. This DKT phenomenon has been widely investigated in the past \cite{Fortes1987}  \cite{Feng1994}   \cite{Hu1996}   \cite{Glowinski1999}   \cite{Uhlmann2005}, either experimentally or numerically, in contrast with the period-doubling cascade and the chaotic dynamics observed by Aidun \& Ding  \cite{Aidun2003} which received less attention.

The goal of this paper is to analyze the dynamics of the doublet for more inertial regimes which had not been investigated so far, and to determine whether the sedimentation process converges to some regular  structure. 
Like in Ref. \cite{Aidun2003}, we focus on the two-dimensional problem, so that particles can be thought of as disks. Throughout the paper, the terms "particles" or "disks" will be used indifferently to refer to these objects.

We are particularly interested in the link between the sedimentation velocity and the volume force driving the motion of the particles. To achieve this, we have developed a numerical algorithm  based on the fictitious domain method (section \ref{MethodeNum}), and used it to simulate the sedimentation of two disks in a vertical channel for $\mbox{Re}_T$ in the range $[2,10]$.  
The control parameter of the various simulations shown below is the non-dimensional apparent weight of the particles (which is the same for both particles since they are identical):
\begin{equation}
F=\frac{\pi}{4} G (\rr-1)
\label{defF}
\end{equation}
where $G$ is the Galileo number ($G={D^3 g}/{\nu ^2}$)
and $\rr$ is the particle / fluid density ratio. \rouge{
To ease the comparison, the same density ratio as  Aidun \& Ding \cite{Aidun2003} will be considered throughout the paper ($\rho_r = 1.002$).  }
Lengths and times have been set non-dimensional by using   $D$ and   $\nu$.
The range of $F$ investigated in Ref. \cite{Aidun2003} corresponds to $F \le 250$ \rouge{(or, equivalently, $G \le 159000$)}. This "weakly inertial" regime will be re-visited in section \ref{Aidunrevisited} below. The new phenomena investigated in the present paper appear when $370 \le F \le 500$
\rouge{(i.e. $236000 \le G \le 318000$)} and will be presented  in section \ref{Fgt250}. \rouge{
These non-dimensional numbers correspond for example to particles of a few $mm$ wide, slightly heavier than the fluid, and sedimenting in water. }
It will be shown that another chaotic attractor exists in this range, and that this attractor affects the sedimentation process. A conclusion will be drawn in section \ref{concl}.

%%%%%%%%%%%%%%%%%%%%%%%%%%%%%%%%%%
\section{Problem description and numerical approach}
\label{MethodeNum}
  
  We consider a pair of circular disks with a diameter $D$, settling in a vertical plane channel filled with a Newtonian fluid, with viscosity $\nu$. The length of the channel is infinite in the vertical direction, and its width is $W = 4 D$. 
A Direct Forcing Fictitious Domain method  \cite{Yu2007}  has been developed to solve the motion equation of both the fluid and the particles. It consists in solving the Navier-Stokes equation over a fictitious domain $\Omega$ including the fluid and the particles, which will be denoted as $P_1(t)$ and $P_2(t)$ in the following. A volume force $\lambda$ is then applied within the disks to impose a rigid-body motion. The resulting motion equations are:
\begin{eqnarray}
\label{ns1}
{\rho_f} \left(\frac{\partial \textbf{u}}{\partial t}+ (\textbf{u}.\boldsymbol{\nabla})\textbf{u} \right) &=& - \boldsymbol{\nabla}p + \mu \lapu + \boldsymbol{\lambda} 
\quad \textnormal{	in } \quad {\Omega},\\
\label{ns2}
{\divu}&=& 0  
\quad \textnormal{	in } \quad {\Omega},\\
 \textbf{u} &=& \textbf{U} _i + \boldsymbol{\omega} _i \times \textbf{r} 
 \quad \textnormal{	in } \quad {P_i(t)},
 \label{CL}
\end{eqnarray}
where $\textbf{u}$ is the velocity field, $p$ is the pressure, $\rho_f$ is the fluid density, $\mu$ is the dynamic viscosity,
and ($\textbf{U}_i$,$\boldsymbol{\omega}_i$) denote the translational and rotational velocities of particle $P_i$ respectively. The linear and angular momentum equation read:
\begin{eqnarray}
\label{eqU}
(1-\frac{1} {{\rho_r} }) m_i ({\frac{\ud \textbf{U} _i}{\ud t}}-\textbf{g}) &=& - \int_{P_i(t)} \boldsymbol{\lambda} \, \ud S , \\
(1-\frac{1} {{\rho_r} }) \frac{\ud ({\textbf{J} _i \boldsymbol{\omega} _i})}{\ud t} &=& - \int_{P_i(t)} \textbf{r} \times \boldsymbol{\lambda} \, \ud S,
\label{eqw}
\end{eqnarray}
where $m_i$ and $\textbf{J} _i$ denote the mass and moment of inertia tensor respectively.

\begin{figure}[h] 
\center{\includegraphics[scale=0.9]{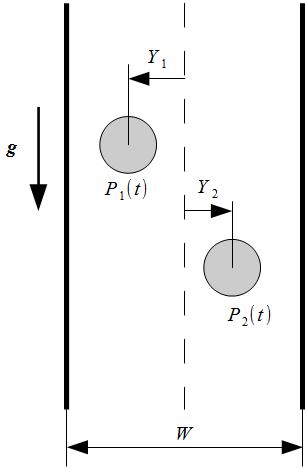}}
\caption{Sketch of the particle pair in a confined domain.  }
\label{schem2part}
\end{figure}

Eqs.\ (\ref{ns1})-(\ref{ns2})-(\ref{CL}) are discretized by means of a Finite Differences algorithm on a cartesian staggered mesh\cite{Hofler2000}, and by using a projection method (Chorin \cite{Chorin1968}  ; Temam  \cite{Temam1969}). Time-stepping is done by means of a 2nd-order Adams-Bashforth scheme for advection, together with a 2nd-order Crank-Nicolson scheme for the diffusion term. Eqs.\ (\ref{eqU}) and (\ref{eqw}) are discretized by means of a Collocation Points Method  \cite{Yu2007}. The resulting discretized equations, in non-dimensional form, read: 
$$
\frac{\ustar - \upn}{\disct} = - \boldsymbol{\nabla} p^n + \frac{1}{2} (3 (\textbf{u}.\boldsymbol{\nabla}\textbf{u})^{n} - (\textbf{u}.\boldsymbol{\nabla}\textbf{u})^{n-1})  
$$
 \begin{equation}
+\frac{1}{2} ( {\boldsymbol{\nabla}}^2 \upn +  {\boldsymbol{\nabla}}^2 \ustar ) + \lp ,
\end{equation}
with:
\begin{eqnarray}
\textbf{u}^{n+1} &=& \textbf{U} _i^{n+1} + \boldsymbol{\omega} _i^{n+1} \times \textbf{r},\\
\boldsymbol{\nabla} . \ustar &=& 0,
\label{mr}
\end{eqnarray}
where $\mathbf{u}^*$ is a provisional velocity which, in general, does not satisfy the rigid-body motion within particles. The translational and angular velocities of each particle satisfy:
\begin{eqnarray}
 \label{nfd}
(1-\frac{1} {{\rr} }) v ({\frac{\ud \textbf{U} _i^{n+1}}{\ud t}}- G \, {\hat {\textbf{g}}}) &=& - \int_{P_i(t)} \negmedspace\negmedspace\negmedspace\negmedspace 
 \boldsymbol{\lambda}^{n+1}  \ud S ,\\
\label{ntd}
(1-\frac{1} {{\rr} }) \frac{\ud ({\textbf{J} _i \boldsymbol{\omega} _i}^{n+1})}{\ud t} &=&- \int_{P_i(t)}\negmedspace\negmedspace\negmedspace \negmedspace\negmedspace\negmedspace 
\textbf{r} \times \boldsymbol{\lambda}^{n+1}  \ud S,
\end{eqnarray}
where ${\hat {\textbf{g}}} $ is the unit vector in the direction of gravity and 
$v$ is the particle volume.
The volume force is then chosen to impose a rigid-body motion within $P_1(t)$ and $P_2(t)$ (see also Yu \& Shao  \cite{Yu2007}):
\begin{equation}
 \label{lambdaa}
\frac{\upp -\ustar}{\disct}=\lpp -\lp.
\end{equation}
\rouge{
The method has been validated on a number of benchmarks involving either isolated or interacting two-dimensional particles at Reynolds numbers varying between 0.1 and a few hundred. It has been systematically compared to existing results from the literature: fixed or oscillating cylinder \citep{Happel1965,Harper1967,Park1998,Uhlmann2005}  ; unique cylindrical particle sedimenting in a vertical channel with either axial or  asymmetric initial position    \citep{Wachs2009,Uhlmann2005} ; pairs of particles interacting in a vertical channel \citep{Feng1994,   Wachs2009,  Uhlmann2005}.  These benchmarks are presented in details in Ref. \cite{Verjus2015}. 
}
In the following section, most of the results by Aidun \& Ding  \cite{Aidun2003}, which had been obtain by using a completely different method (Lattice Boltzmann), will be revisited and confirmed.

%%%%%%%%%%%%%%%%%%%%%%%%%%%%%%%%%%%%%%
\section{$F \le 250$:  oblique doublet  and subharmonic cascade } 
\label{Aidunrevisited}

\begin{figure}[h] 
\center{\includegraphics[scale=0.6]{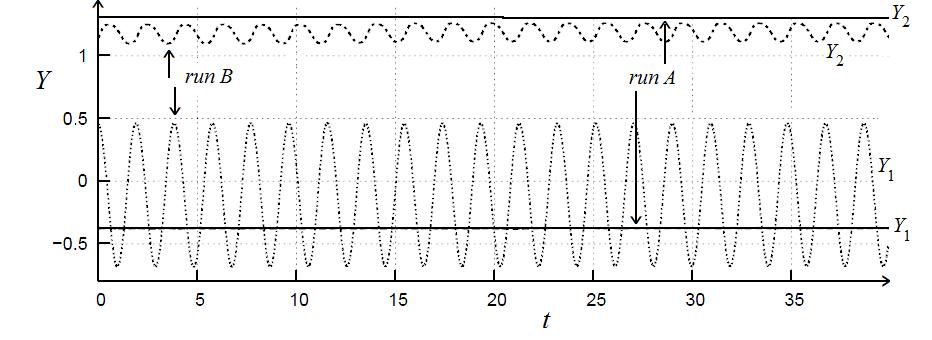}}
\caption{Evolution of the horizontal coordinate of the centers of the particles when $F = 136.77$. Solid lines correspond to the steady oblique doublet ({\it run A}, lower branch of Fig.\ \ref{hyst}), and  dashed lines correspond to  periodic oblique doublet ({\it run B}, upper branch of Fig.\ \ref{hyst}). }
\label{DoubletsObliques}
\end{figure}

\begin{figure}
\center{\includegraphics[width= .6\textwidth]{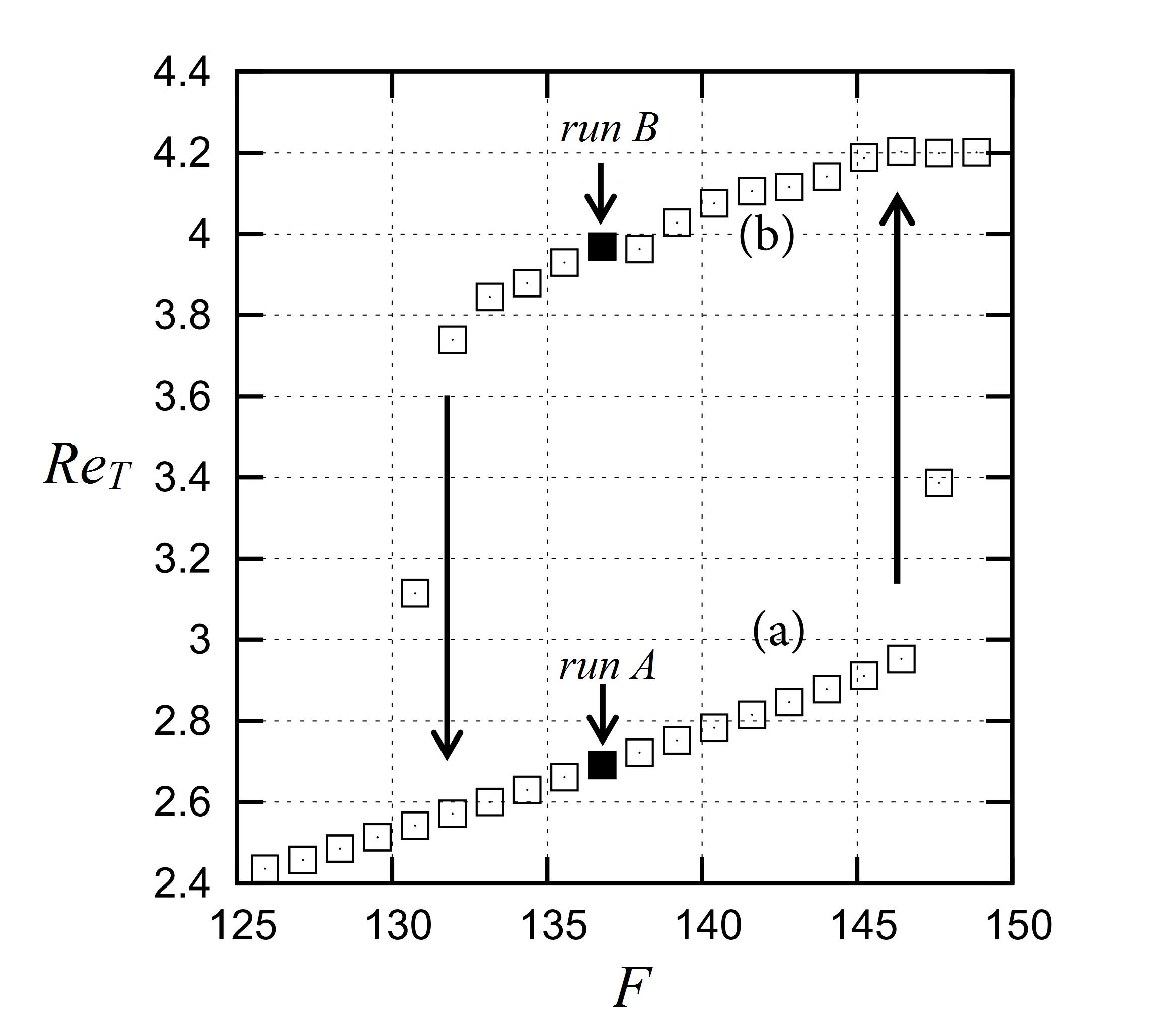}}
\caption{Terminal Reynolds number vs. the non-dimensional driving force $F$  in the range [125,150].
The lower branch  (a) corresponds to a slow sedimentation in the form of a steady oblique doublet, whereas the upper one ("fast" branch, (b))  is the time-periodic doublet. Runs $A$ and $B$ of Fig.\ \ref{DoubletsObliques} are marked with a black square. }
\label{hyst}
\end{figure}

Fig.\ \ref{DoubletsObliques} shows the evolution of the horizontal coordinates $Y_1$ and $Y_2$ of the centers of both particles,
 when  $F = 136.77$. Two runs are shown, corresponding to two initial positions of the particles. 
 In one case ({\it run A}, solid lines),  \rouge{the pair is initially released in a  horizontal position with $Y_1=-0.4$ and $Y_2=1.4$}, and  converges to a steady oblique doublet. In the second case ({\it run B}, dashed lines) \rouge{the pair is also released in a horizontal position, but nearer to the axis with $Y_1=-0.1$ and $Y_2=1.0$}, and reaches a periodic regime.  
 The terminal Reynolds number $\mbox{Re}_T$ of the pairs is very different in both cases. It is plotted in Fig.\ \ref{hyst},  for $F$ varying in the range [125,150]. Two branches co-exist: the lower one ("slow" branch, (a))  is the steady oblique doublet, whereas the upper one ("fast" branch, (b))  is the time-periodic doublet. 
Below $F=130$, only the steady branch exists. A Hopf bifurcation  takes place at $F\simeq 131$.
These results confirm previous analyses  \cite{Aidun2003} \cite{Feng1995}, except that our lower branch clearly corresponds to a steady doublet, in agreement with Feng {\it et al.}   \cite{Feng1995}, but in contrast with  Aidun \& Ding  \cite{Aidun2003} who observe a periodic doublet there. This might be attributed  to a numerical artefact, since the amplitude of the oscillations of the lower branch of
Aidun \& Ding is very small ($0.025D$), and  of the order of their mesh size ($0.03125D$).  By using a finer mesh our computations always led to a steady oblique doublet on the lower branch, like the one shown by the solid line of Fig.\ \ref{DoubletsObliques}.

\vskip.3cm

\begin{figure}[h]  
\hspace*{-.6cm}\includegraphics[width= .85\textwidth]{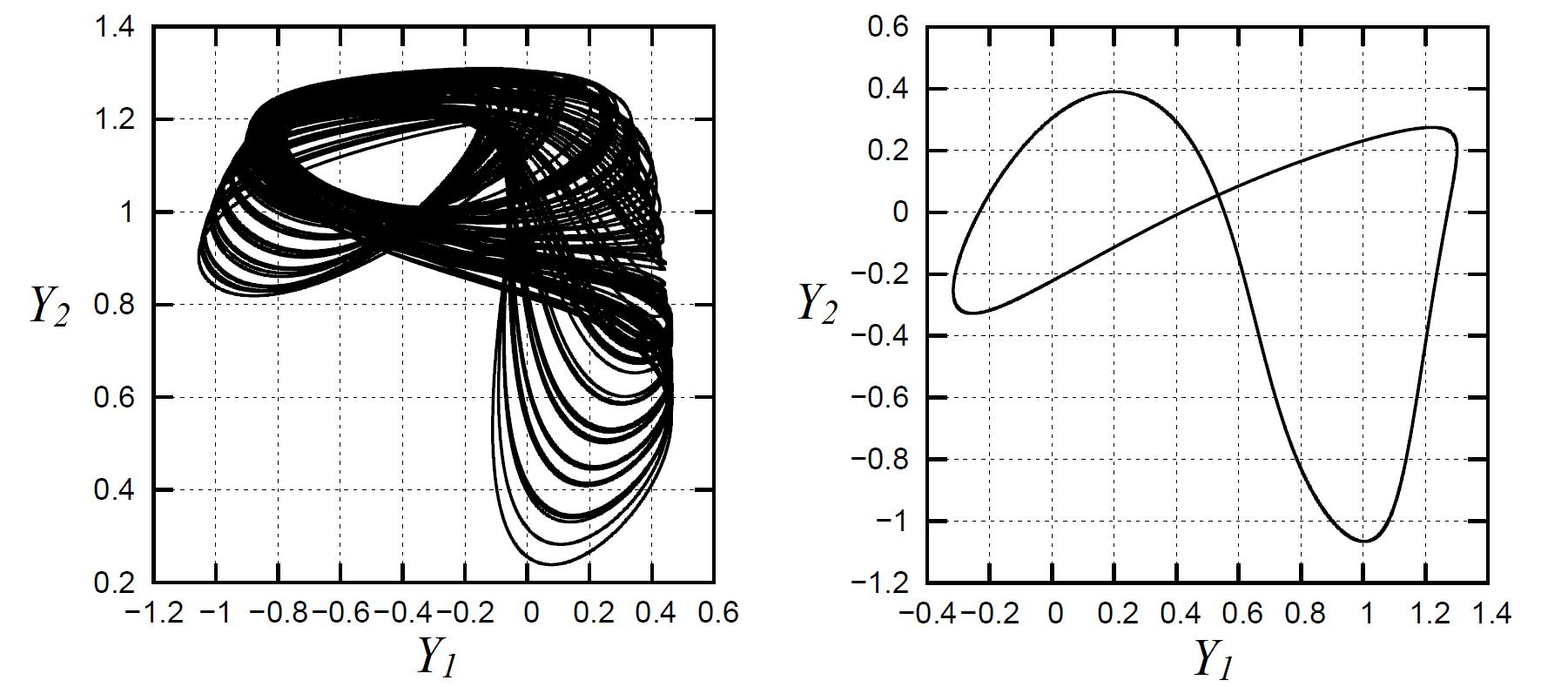}
\caption{Left: chaotic attractor resulting from the subharmonic cascade, $F=197.68$. Right: return to periodicity (DKT dynamics), $F=241.61$.}
\label{DKTWSD4}
\end{figure}

The lower branch of  Fig.\ \ref{hyst} can be approximated by the affine formula :
$
\mbox{Re}_T=0.0251 F-0.738$,
and the upper branch by:
$
\mbox{Re}_T=0.030 F-0.101  $.
Both branches are stable and co-exist when $130.74 \le F \le 145$. However, our simulations show that only the upper one persists, with a well-defined period $T$, when $F$ is above this range. In addition, increasing $F$  above 145 leads to a series of period-doubling bifurcations. The first one appears when $F \simeq 146$, and the period of the doublet jumps from $T$ to $2T$.    The $2T \to 4T$ bifurcation occurs
when  $F \simeq 156$, and $4T \to 8T$  occurs when $F \simeq 158$.
 Fig.\ \ref{DKTWSD4} (left) shows the dynamics in the plane ($Y_1$, $Y_2$) after a large number of period-doubling bifurcations: a chaotic attractor, already observed by Aidun \& Ding   \cite{Aidun2003} appears. Increasing again the non-dimensional weight leads to a more regular, periodic,
dynamics (Fig.\ \ref{DKTWSD4} (right)). It corresponds to the DKT regime discovered by Feng {\it et al.}  \cite{Feng1995}. In the next section we focus on this inertial regime.

    \rouge{
    Note that the oblique doublet observed here is often explained by invoking the role of the
intrinsic rotation of the particles, which would create a Magnus lift force. This force is opposed to the repulsive force induced by the vertical walls, and the competition between these two effects would maintain the stability of the doublet. If this explanation were relevant, one could argue that, in the absence of rotation, the sedimentation structure should be quite different. To check this point we have done a series of runs with exactly the same physical conditions as the ones used in  section \ref{Aidunrevisited}, but by removing the rotational degree-of-freedom of the objects. No significant difference emerged, and the dynamics of the oblique doublet was only slightly perturbed. Indeed, the lift force does not necessarily requires the particles to rotate around their centers, since this force results from the asymmetry of the disturbance flow due to the inclusion, which can exist even if the particle does not rotate.  
    }

%%%%%%%%%%%%%%%%%%%%%%%%%%%%%%%%%%%%%%
\section{$F \ge 250$:  horizontal doublet and quasi-periodic route to chaos}
\label{Fgt250}

\begin{figure}[h] 
\center{\includegraphics[width= .45\textwidth]{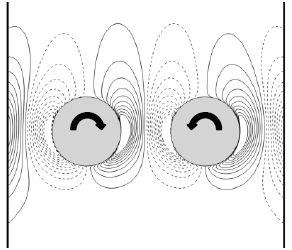}}
\caption{Steady symmetric horizontal doublet at $F=400.12$. Solid and dashed lines correspond to positive and negative vorticity contours respectively. In this configuration, the effective blockage ratio \rouge{(defined as the apparent total diameter of the pair divided by the gap of the channel) is maximum}.}
\label{horizdoublet}
\end{figure}

\begin{figure}[h] 
\center{\includegraphics[width= .9\textwidth]{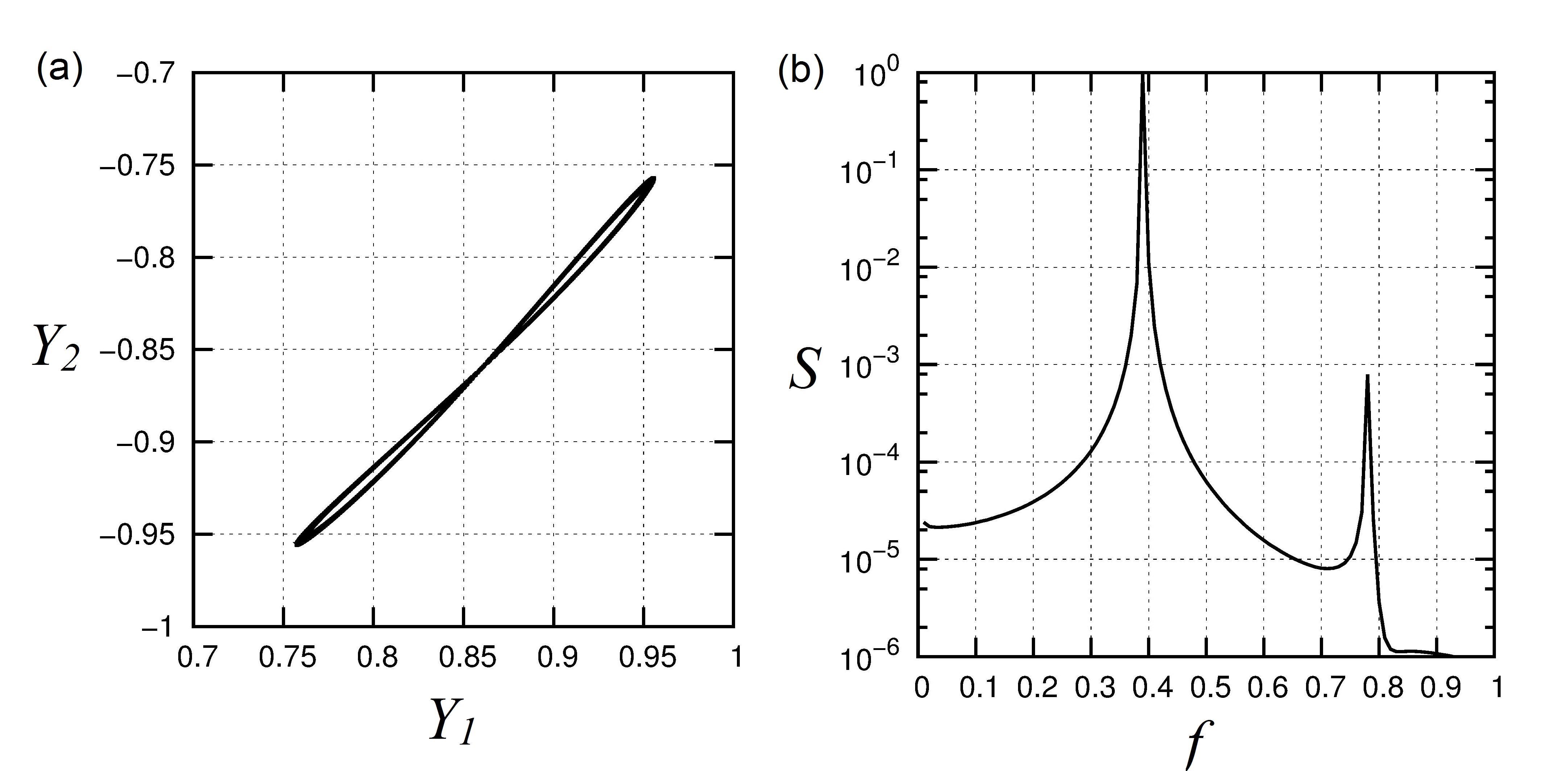}}
\caption{Periodic oscillations of the horizontal doublet when \rouge{$F=427$}.  (a): dynamics in the $(Y_1,Y_2)$ plane. (b): Power spectrum $S(f)$ of $Y_1(t)$ showing peaks at \rouge{$f_1 \simeq 0.397 \, Hz$ and $2 f_1$}.  }
\label{T1}
\end{figure}

\begin{figure}[h] 
\center{\includegraphics[width= .9\textwidth]{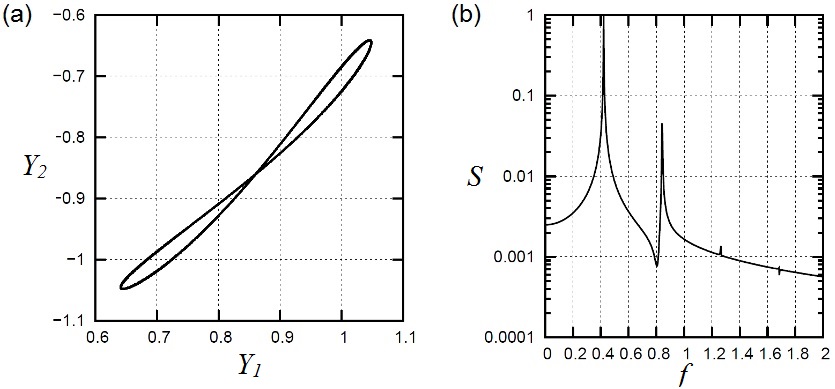}}
\caption{Periodic regime at $F=480$. (a): dynamics in the $(Y_1,Y_2)$ plane.  The power spectrum (b) shows harmonics of the fundamental frequency $f_1 \simeq 0.421 \, Hz$. }
\label{T1bis}
\end{figure}

We now focus on regimes which had not been explored in details so far.
The usual DKT regime is observed to persist when $F>250$, up to $F \simeq 400$.
When $F$ approaches this value, we observe that the DKT phenomenon co-exists with a  steady horizontal structure, the vorticity field of which is shown in Fig.\ \ref{horizdoublet}. To our knowledge, this structure had not been observed in previous analyses.
According to the initial orientation of the pair, particles either perform DKT, or converge to a horizontal structure as the one shown in Fig.\ \ref{horizdoublet}.
In this case the flow is highly symmetric with respect to the middle-line, and the trajectory of the particle centers is perfectly vertical. Both objects rotate as if rolling on the nearest wall. When $F > 400$ the DKT no longer exists, but the quasi-horizontal structure persists. It is observed to exist irrespective of  the disks' initial positions. A Hopf bifurcation appears at $F \simeq 400$, and the horizontal structure becomes time-periodic:  Fig.\ \ref{T1} shows this oscillation in the $(Y_1,Y_2)$ plane, as well as the power spectrum of $Y_1(t)$,  when $F=427$. A frequency $f_1 \simeq 0.397 \, Hz$ and its harmonic $2 f_1$ is clearly visible. In this regime, the doublet remains perfectly horizontal, the particles sediment with a zigzag motion while keeping their distance constant.
Particles never cross the axis of the channel, as $Y_1$ and $Y_2$ remain positive and negative for all times, respectively.
 When $F$ increases, the regime remains periodic until $F \simeq 486$.  Fig.\ \ref{T1bis} shows the case $F=480$. A  fundamental frequency 
$f_1 \simeq 0.421 \, Hz$ and its harmonics are visible. \rouge{Particles overtake each other periodically, i.e. the difference between their vertical coordinates ($X_1-X_2$) changes sign periodically. }

\begin{figure}[h] 
\center{\includegraphics[width= .9\textwidth]{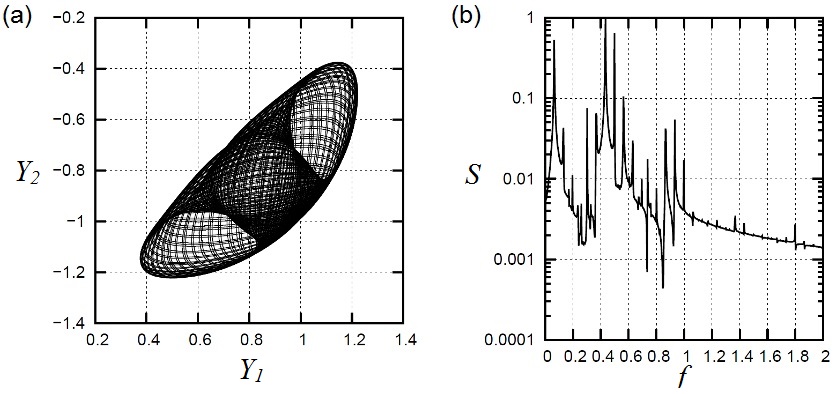}}
\caption{Quasi-periodic oscillations of the horizontal doublet when $F=489.5$. (a): dynamics in the $(Y_1,Y_2)$ plane. (b): power spectrum of $Y_1(t)$.   }
\label{T1T2}
\end{figure}

 \pagebreak
 
\begin{figure}[h] 
\center{\includegraphics[width= .8\textwidth]{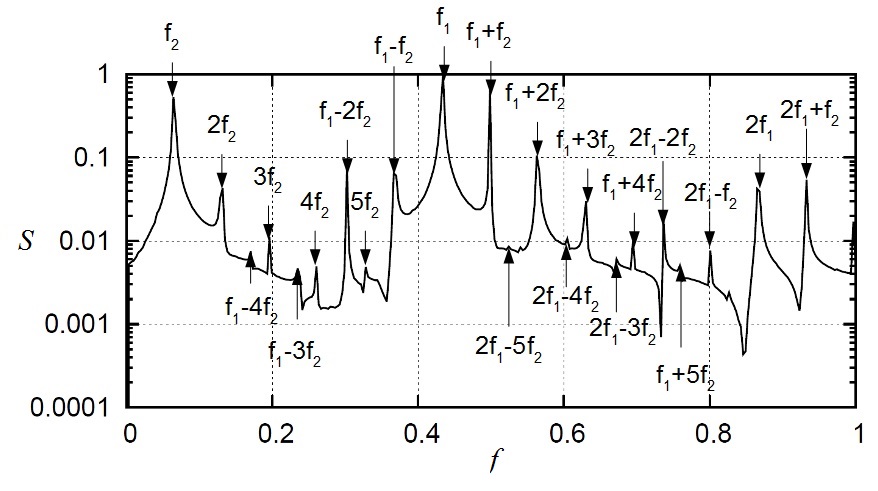}}
\caption{Magnification of the spectrum of Fig.\ \ref{T1T2}.  Fundamental frequencies are $f_1 \simeq 0.434 \, Hz$ and $f_2 \simeq 0.064 \, Hz$. }
\label{CombiLinf1f2}
\end{figure}

Under increasing $F$, the phase portrait becomes more complex  (Fig.\ \ref{T1T2}(a), $F=489.5$).  The power spectrum of $Y_1(t)$ shows a large number of peaks (Fig.\ \ref{T1T2}(b)), corresponding to linear combinations of two fundamental frequencies $f_1 \simeq 0.434 \, Hz$ and $f_2 \simeq 0.064 \, Hz$ (Fig.\ \ref{CombiLinf1f2}). Particles still overtake each other unceasingly, but in a non-straightforward manner. The trailing particle rotates
for some time while remaining behind the leading one, then overtakes it.  
Note that particles do not cross the axis of the channel \rouge{(i.e.   $Y_1(t) > 0$ and $Y_2(t) < 0$ for all $t$)}, like in the periodic cases above.
  
\begin{figure}[h] 
\center{\includegraphics[width= .8\textwidth]{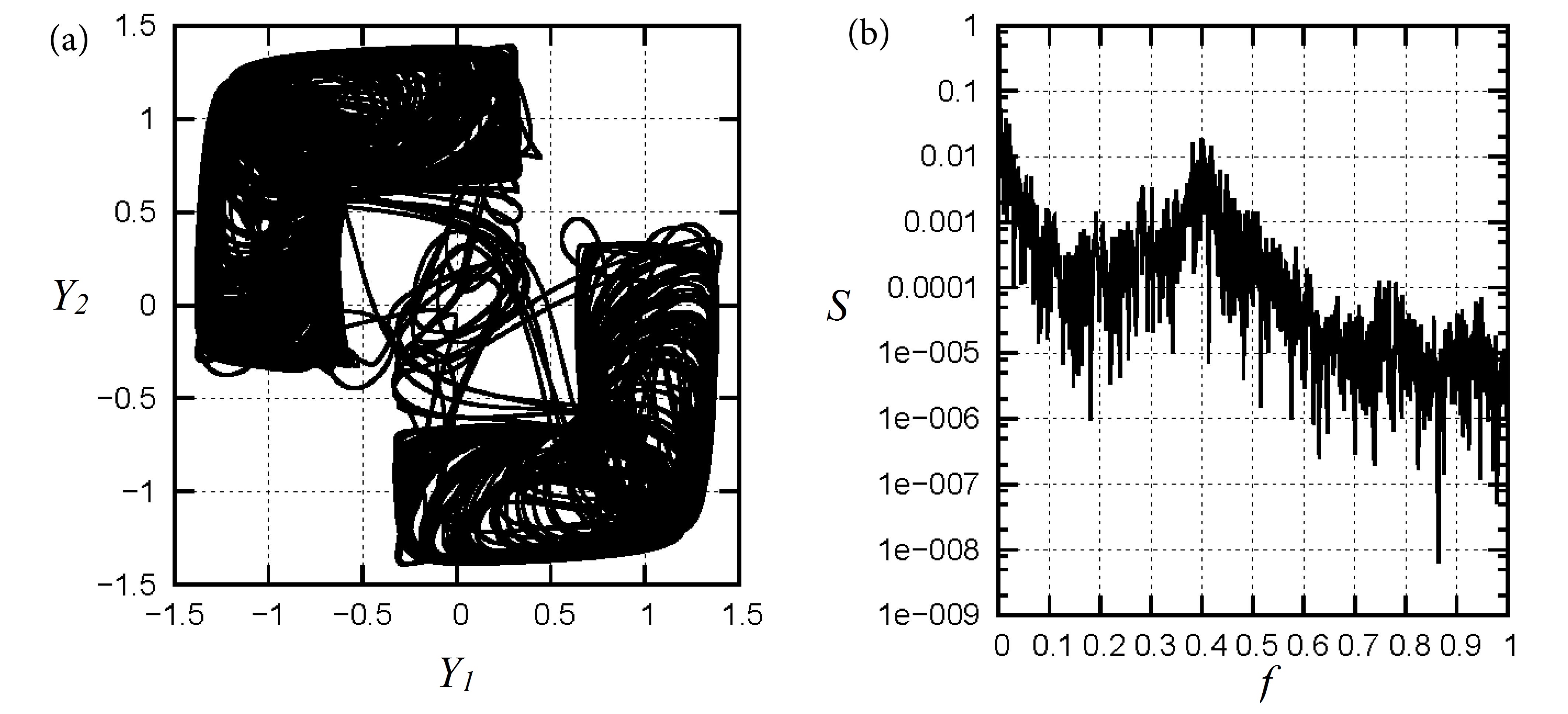}}
\caption{Chaotic attractor (a) and corresponding spectrum (b) when $F = 527.78$.  }
\label{T1T2T3chaos}
\end{figure}

When $F$ is above 507, a chaotic dynamics takes place. Fig.\ \ref{T1T2T3chaos} shows our results when $F=527.78$.  The phase portrait is characterized by very disordered trajectories in a bounded volume of the phase space, where particles cross the channel axis \rouge{(i.e. the sign of $Y_i(t)$ is no longer constant)} in an intermittent manner (Fig.\ \ref{T1T2T3chaos} (a)). The spectrum shows a wide range of frequencies, with a broadband noise structure (Fig.\ \ref{T1T2T3chaos} (b)). This suggests that the Ruelle-Takens scenario is taking place here, and that chaos occurs after the appearance of a third frequency.


~ 

\newpage

\section{Global bifurcation diagram}

\begin{figure}[h] 
\center{\includegraphics[width= .99\textwidth]{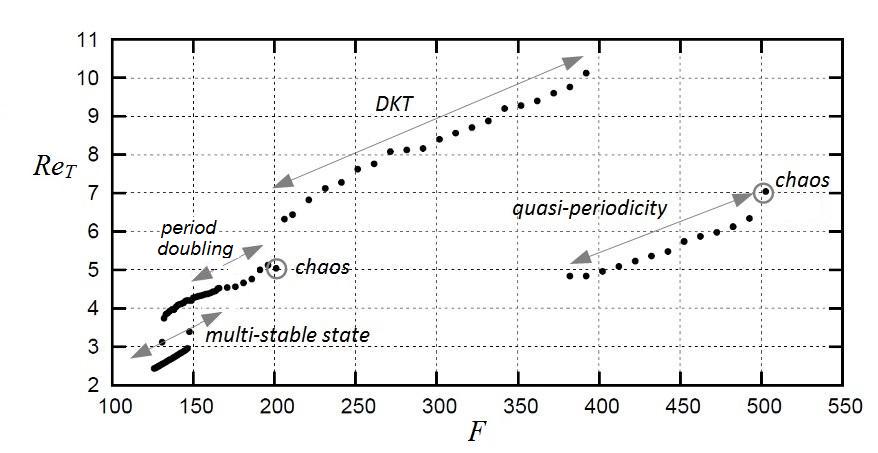}}
\caption{Global diagram:  settling Reynolds number $\mbox{Re}_T$ versus the non-dimensional driving force $F$.  }
\label{GlobalDiag}
\end{figure}

These results show that the settling of the particle pair is a very complex phenomenon, even in this simple two-dimensional configuration. In particular, the terminal Reynolds number of the pair is difficult to predict. 
To give a general view of the sedimentation process, 
we summarize the  various regimes described above by  plotting the terminal Reynolds number $\mbox{Re}_T$ versus $F$ (Fig.\ \ref{GlobalDiag}), up to $F=500$. We observe that $\mbox{Re}_T$ is a piecewise increasing function of $F$, \rouge{containing quasi-discontinuities at bifurcation points where $\mbox{Re}_T$ decays or increases abruptly}. Therefore, increasing the non-dimensional weight $F$ will not lead systematically to higher settling Reynolds numbers. 
The most spectacular drop of settling velocity appears when the periodic DKT becomes unstable, and the quasi horizontal stable structure appears. This happens when $F$ increases from 370 to 400. 
This abrupt decay is related to the abrupt change in the {\it effective} blockage ratio (defined here as the apparent total diameter of the pair divided by the gap of the channel). It is minimum, and equal to $D/W$, when the doublet is in a vertical  position (i.e. the leading particle hides entirely the trailing one). In contrast, it is maximum and equal to $2D/W$ when the doublet is in the horizontal position.

The decay of settling velocity is very remarkable here, as it is mostly divided by two when $F$ varies from 370 to 400. However, this effect is rather common in that it can be observed in many elementary dynamical systems. For example, consider a two-dimensional system $(u(t),y(t))$ such that
$
\dot u = F - (1+y^2) u.
$
It corresponds to a forced system (with a constant driving force $F>0$) submitted to a friction force with friction coefficient $1+y^2$ depending on the second variable $y(t)$. (The variable $u$ can be thought of as the settling velocity of the system of particles, whereas $y$ plays the role of the effective blockage ratio which increases the efficiency of viscous friction.) Suppose that, for $F < F_0$, say, the system has a stable equilibrium position at $y=y_a = 0$ and  $u=u_a = F/(1+y_a^{2}) = F$. Therefore, one can expect the system to remain in the vicinity of this state for long times, provided the initial conditions have been chosen close enough to $(u_a,y_a)$. Now, suppose that the position $y_a =0$ is no longer stable for $F>F_0$, and that a new asymptotically stable position appears there, e.g. $y_b=1$ (which would correspond to the horizontal doublet in our simulations). Then the system will quit the vicinity of the terminal velocity $u_a = F$, and is likely to converge to the vicinity of a much smaller terminal velocity, that is $u_b = F/(1+y_b^2) = F/2$. 
Let $\langle u \rangle$ denote the velocity $u$ averaged over long times. Assuming that $\langle u \rangle$ is close to the stable equilibrium solution $u_a$ or $u_b$, one can expect that the graph of $\langle u \rangle$ versus $F$ will display a steep decrease at $F=F_0$, like the one observed in Fig.\ \ref{GlobalDiag}. 
This behavior corresponds to a kind of "obstruction effect", in that increasing the cause of the motion ($F$) leads to a larger effective blockage ratio, and to a slower motion. In particular, even if both $u_a$ and $u_b$ increase separately with $F$, the  {\it stable} equilibrium solution $\{u_a$ or $u_b\}$, and therefore $\langle u \rangle,$ is only piecewise increasing, and, in the present case, decays with $F$ in the vicinity of $F_0$. 
%
%We have tested these conjectures by solving Eq.\ (\ref{u}), together with:
%\begin{equation}
%\ddot y = (F-F_0) (y-y^2),
%\label{y}
%\end{equation}
% which is the simplest dynamical equation providing a linearly stable equilibrium point $y_a=0$ for $F<F_0$ (degenerate center point), which becomes unstable when $F > F_0$. Conversely, the equilibrium point $y_b=1$ is stable only when $F > F_0$. We have set $F_0=1$, and solved Eqs.\ (\ref{u}) and (\ref{y}) numerically with an initial condition $u(0)=0$ and $y(0)$ chosen arbitrarily between $y_a$ and $y_b$. We observe that the long-time average velocity  $\langle u \rangle$ remains close to $u_a=F$ when $F < F_0$ (see Fig.\ ), then abruptly jumps down to $u_b=F/2$ as soon as $F > F_0$. 

Finally, note that all the initial positions used in the present  simulations for $F > 400$ lead to the quasi-horizontal doublet structure. We have checked however that particles released in a perfectly vertical manner ($Y_1(0)=Y_2(0) = 0$), settled by conserving their initial vertical orientation, and acquired  a terminal velocity which is much larger than the one of the quasi-horizontal doublet. Similarly, particles injected symmetrically along a horizontal line ($X_1(0) = X_2(0)$ and $Y_1(0) = - Y_2(0)$) will  keep their horizontal orientation, and sediment slowly. However, in both cases, any small disturbance forces the particles to join the quasi-horizontal slow structure. This suggests that this structure is asymptotically stable (attracting) and has a large basin of attraction.

\begin{figure}[h] 
\center{\includegraphics[width= .99\textwidth]{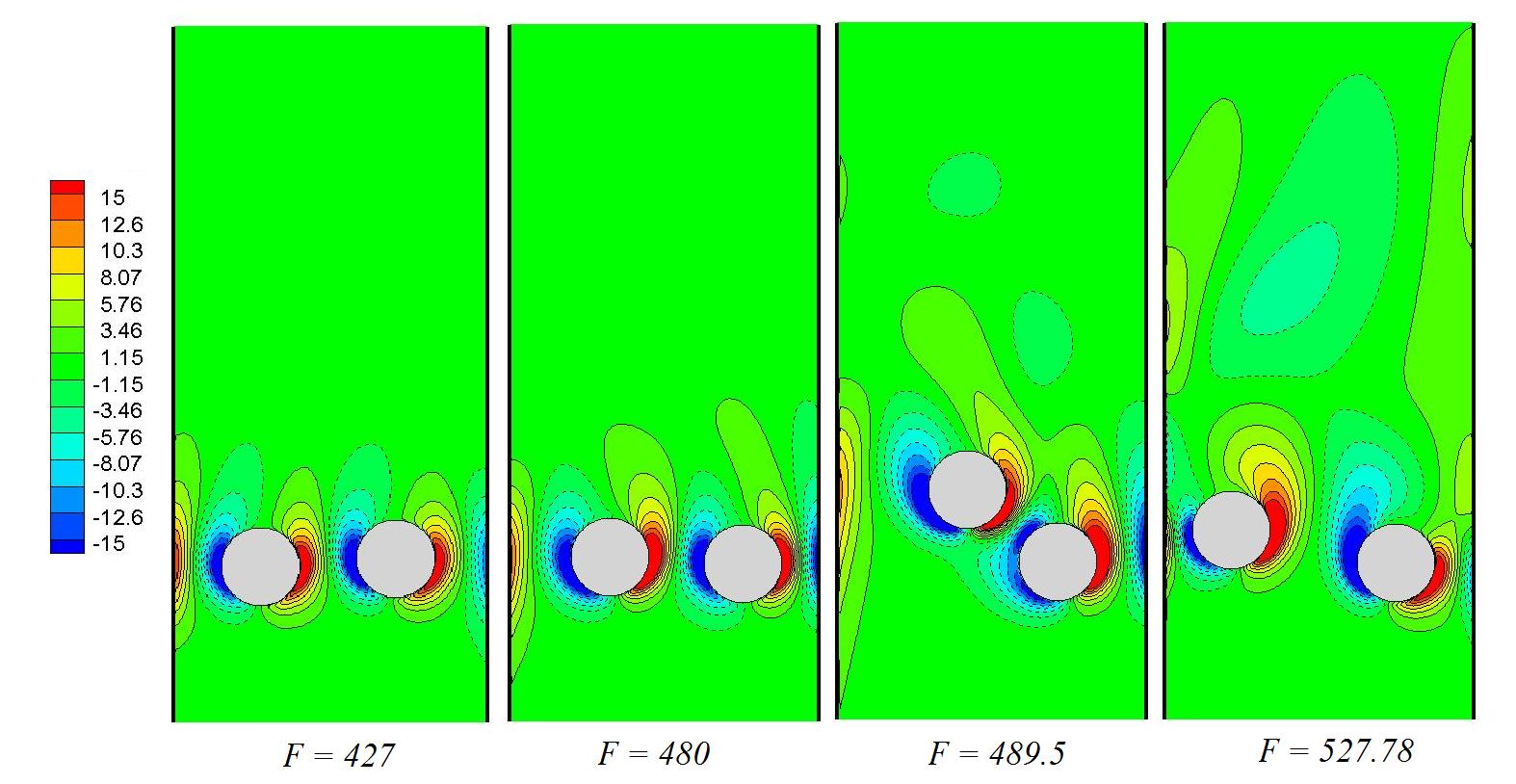}}
\caption{Instantaneous vorticity fields for increasing driving forces : periodic regimes ($F=427$ and $F=480$), quasi-periodic ($F=489$) and chaotic regime ($F=527$).  }
\label{Fields}
\end{figure}

\rouge{
The various regimes described above correspond to a complex interaction between the particles and the fluid, as shown in the instantaneous vorticity fields of Fig.\ \ref{Fields}, corresponding to the very same parameters as the ones of Figs.\ \ref{T1}, \ref{T1bis}, \ref{T1T2} and \ref{T1T2T3chaos}. In the periodic regime observed at $F=427$, the wake of the objects is highly localized. When the driving force increases, the wake looses its symmetry and becomes more unsteady. The vorticity produced by particles creates a non-trivial flow, showing evidence of vortex shedding, which in turns affects the particles. The disordered flow structure is even more pronounced in the chaotic case $F=527.78$. This suggests that a strong coupling exists between flow and particles, and that any attempt to reduce the dynamics should take into account not only the degrees-of-freedom of the particles, but also the effective degrees-of-freedom of the flow.
}

\section{Conclusion}
\label{concl}

The computations presented in this paper reveal some new features of the settling of inertial disks in a confined domain. The multiple stable-state, the hysteresis and the chaotic attractor discovered by Aidun \& Ding   \cite{Aidun2003} have been recovered when the non-dimensional driving force $F$ is below 200. Our results are in quantitative agreement with the ones of these authors, except that the lower branch of the bifurcation diagram of Fig.\ \ref{hyst} corresponds to a steady position in our case. We attribute these differences to some numerical artefact.

For  more inertial regimes, the collisional  DKT phenomenon takes place, in agreement with previous analyses. The particle Reynolds number increases almost monotonically with $F$ over the range $200 \lesssim F \lesssim 400$. An abrupt change occurs for larger $F$: the DKT vanishes and particles tend to join a non-collisional attracting quasi-horizontal structure, corresponding to a slow sedimentation. 
Settling, though slow, can then be  steady, periodic, quasi-periodic or chaotic, according to the values of $F$. A transition towards chaos occurs on this branch as $F$ increases, but the route leading to chaos is different from the subharmonic cascade observed by Aidun \& Ding \cite{Aidun2003} at smaller $F$. Indeed, a quasiperiodic route is  observed here when  $400 \lesssim F \lesssim 500$, leading to a chaotic attractor where particles cross the axis of the channel in an intermittent manner. 

 The link between the (non-dimensional) particle weight $F$ and the (non-dimensional) settling velocity 
 $\mbox{Re}_T$ is therefore extremely complex in this apparently simple situation.   The most spectacular effect is the abrupt decrease in settling velocity at $F \simeq 400$: here,  the doublet takes the form of a nearly horizontal structure, the channel is therefore partially "blocked" by particles, and sedimentation is slow. In a sense, such a behaviour shares some common features with the Braess effect observed in electrical circuits or in pipe flows\cite{pala2012transport}
 \cite{cohen1991paradoxical}: near some critical value, increasing the cause of the motion will lead to a larger blockage and to a slower motion.

Finally, to understand further the complex dynamics involved in this sedimentation process we intend to perform simulations with larger driving forces. Also, three dimensional situations, that is pairs of spheres settling in a vertical channel, will be considered in the near future.
 
 \vskip.3cm
 \noindent
{\it This paper is dedicated to the memory of Professor Alexander Ezersky.}

% \bibliography{global}
 
 %%%%%%%%%%%%%%%%

\end{document}